# Preparation and properties of Er–doped ZrO$_2$ nanocrystalline phase-separated preforms of optical fibres by MCVD process

Anirban Dhar[a*], Ivan Kasik[a], Bernard Dussardier[b], Ondrej Podrazky[a] and Vlastimil Matejec[a]

[a]Institute of Photonics and ElectronicsASCR, v.v.i , Chaberska 57, 18251 Prague-8-Kobylisy, Czech Republic

[b]Laboratoire de Physique de la Matière Condensée, Université de Nice-Sophia-Antipolis, CNRS UMR 6622, Avenue Joseph Vallot, Parc Valrose, 06108 NICE CEDEX 2, France

dhar@ufe.cz

**Abstract:** The fabrication of Er-doped ZrO$_2$-based nanocrystalline phase-separated silica optical preforms by the MCVD and solution doping techniques is presented. Fabricated preform cores are nearly transparent and contain phase-separated rare-earth doped nanocrystalline particles with diameters mainly in a range from 20 to 80 nm. High concentrations of erbium and aluminium in preform cores of about 0.3 and 14 mol%, respectively have been achieved without defects on the core-cladding interface. Spectral losses in a range 800-1600 nm and fluorescence spectra of erbium ions around 1550 nm measured on a fibre drawn from the preform are reported.

## 1. Introduction

Rare-earth (RE) doped optical fibres have been developed and successfully implemented in optical amplifiers and fibre lasers since late 1980. Among the different rare-earth ions; erbium (Er) has proved itself as one of significant candidates for applications in erbium-doped fibre amplifiers (EDFAs), one of revolutionary discoveries in the field of fibre optic telecommunication [1]. Since the discovery of the EDFAs, an extensive research has been carried out to improve spectroscopic properties of fabricated erbium-doped fibres (EDFs) in order to broaden the amplification bandwidth.

For achieving the fluorescence enhancement as high as possible different host materials other than pure silica, e.g. doped silica, phosphate glasses, telluride glasses, fluoride glasses, bismuth-based glasses doped with Er ions have been investigated [2-6]. These different host materials have shown their own advantages (wider emission, higher Er doping level etc.) but considering the practical applications in present telecommunication systems, silica based glass still has cutting edge over the other host materials due to its low loss, ruggedness, compatibility with other system components and low manufacture cost.

Therefore, one of new research directions is focused on fabrication processes of silica-based Er-doped optical fibres having properties of multicomponent glasses in order to control Er ion emission properties around 1.5 $\mu$m, a wavelength region which coincides with the lowest attenuation window of silica glasses. This research is aimed at incorporating of rare-earth-doped nanoparticles in fibres using a reliable and mature technology and taking the advantage of their spectroscopic properties to tune amplifiers characteristics, for example. For this purpose different techniques, such as co-sputtering [7], laser ablation [8], ion-implantation [9], pyrolysis [10], sol-gel technique [11-12] and recently developed direct nanoparticle deposition (DND) technique [13], etc, have been developed.

In spite of interesting achieved properties in fibres, most techniques suffer some drawbacks: low temperature glass fusion, or long and costy process, etc. Thus, it would be a challenging task to fabricate optical fibres with regions containing nanoparticles in rich-silica cores by using the conventional Modified Chemical Vapor Deposition (MCVD) process [14-15] combined with the solution doping technique [16]. Already some attempts have been made to study the properties of Er-doped nano-structured fibres completely produced by MCVD and the so-called solution doping technique [17], containing amorphous calcium-phospho-silicate erbium-doped nanoparticles [18]. Some interesting spectroscopic would also benefit from a nanocrystalline environment for the rare-earth ions.

In order to realize this task and develop so called Er-doped nanocrystalline phase-separated optical preforms for drawing optical fibres, it is necessary to select some dopants which allow nucleation processes and precise nucleation and growth control in the glass. Among different silica dopants available, namely $ZrO_2$ can be selected as host material for creating Er-doped nanocrystaline particles since it has many fold advantages. Particularly, the incorporation of $ZrO_2$ enhances the probability of radiative transitions through reducing glass phonon energy ($\sim$470 cm$^{-1}$), $ZrO_2$ exhibit a boiling point (2300°C) higher than the drawing temperature of silica, has a high refractive index, good optical transparency and what is the most important, in the context of this paper, it can act as a good nucleating agent. Sol-gel process has been successfully used for incorporation of crystalline $ZrO_2$ in silica glass [19].

The motivation of the presented work is based on the background summarized above. In this paper we are presenting results of systematic investigation and optimization of different fabrication stages employed for the preparation of Er-doped $ZrO_2$-based nanocrystalline phase-separated optical preforms and fibres by using the MCVD and solution doping techniques. The paper deals also with detail characterization of preform samples and some of the initial results on fibre performances such as attenuation and fluorescence properties,which are necessary prerequisites for practical applications of the prepared fibres in amplifiers or lasers.

## 2. Experimental

### 2.1. Fabrication of preforms and fibres

The fabrication of an Er-doped $ZrO_2$-based nanocrystalline phase-separated preforms of optical fibres consisted of two main stages, namely of the fabrication of a preform and its subsequent thermal treatment. The preform fabrication process was carried out in a conventional MCVD system (Special Gas Controls, GB). The process commenced with the deposition of 3-5 silica cladding layers onto the inner wall of a high quality silica tube (Suprasil F-300). Then a porous core layer composed either of pure silica or silica slightly doped with $P_2O_5$ was applied over the deposited silica layers by the MCVD process at a suitable temperature. Temperature variations during the porous layer deposition were controlled within ±10°C and monitored by an IR pyrometer synchronously moving with a hydrogen-oxygen burner used for external heating of the rotating silica tube.

The porous layer was then soaked with an aqueous soaking solution containing required raw chemicals for a fixed time span. Aqueous soaking solutions were used because selected raw chemicals, namely $LiNO_3$ (99.995% Merck, Germany) $AlCl_3$ anhydrous (99.99% MaTeck GmbH), $BaCl_2,2H_2O$ (99.999% Aldrich), $ZrOCl_2,xH_2O$ (99.99% Aldrich) and $ErCl_3,6H_2O$ (99.995% Aldrich) are only partially soluble in alcohols like methanol or ethanol. $Li_2O$ doping was carried out since the combination of this oxide with $SiO_2$ and $Al_2O_3$ could enhance the mechanical strength of the prepared glass and additionally increase the thermal mechanical strength and transparency of the glass similar to that of lithium-alumino silicate system [20]. Doping with BaO was employed because this dopant possess good nucleating properties and additionally helps to prevent cracking caused by the volume expansion during $ZrO_2$ structural transformations at the preform thermal-treatment [21]. After soaking the layer was dried in air and dehydrated in presence of chlorine. Subsequently, raw chemicals in the porous layer were oxidized and the layer was sintered to a clear glass layer by gradually increasing the temperature to around 1850°C. Finally, a solid rod, the preform, of diameter~ 10-10.5 mm was obtained through the viscous collapse of the tube with the layers deposited on the inner wall at a temperature above 2000°C.

The fabricated preform was thermally treated at a temperature around 1000°C in order to support nucleation and growth processes in the core. Fibres with diameters of 125±0.5 µm were drawn from thermally-treated preforms at temperatures around 2000 °C by using a conventional drawing tower (Special Gas Controls, GB) equipped with a graphite furnace (Centor, USA). A protective UV-curable acrylate jacket (De-Solite, NL) was applied onto the fibres during drawing.

### 2.2. Characterisation of preforms and fibres

Refractive-index profiles (RIPs) of the whole preforms were evaluated both before and after the thermal treatment using an A2600 preform analyser. Thin polished preform sections of a thickness ~ 1.5 to 2 mm were prepared to analyze core glass compositions and dopants distributions in the preform core using Electron Microprobe Analysis (EMA). Scanning Electron Microscopy (SEM) (LEO-S430i) was used to investigate core-cladding interfaces of the fabricated preforms and presence of nanocrystalline-particle within the preform core region. For SEM analysis a sample about 0.8 mm long double-side-polished preform was prepared which was Ag coated (coating thickness ~200A). Samples for Transmission Electron Microscopy (TEM) analysis were also prepared to identify phase-separation and nanocrystals dimensions. These samples were prepared using two different approaches to identify the generation of nanocrystaline particles, their shapes and dimensions. In the first approach we polished a preform sample to a thickness of ~ 0.8 mm and then used Ar-ion mill thinning (~ to 10 µm) to achieve electronic transparency necessary for testing under High Resolution TEM. In the second approach chemical etching was carried out to remove the preform cladding and the extracted core (thickness ~ 1mm) was then crushed to powder. This powder was dispersed in acetone and applied onto a Cu saver for TEM tests. To determine the composition of phase-separated region and nanocrystalline particle formed the electron beam was focused on the selected spot over the investigated region, when the energy dispersive X-ray diffraction (EDX) data were collected.

RIPs and fibre dimensions were measured using an S14 fibre analyser (York Technology, GB). Optical losses of the fibres in a range from 800 to 1600 nm were determined employing the cut-back technique [22]. A white light source was filtered by a computer controlled monochromator and coupled into a segment of the fibre under test. The output optical power $P_1(\lambda)$ from the fibre was measured, where $\lambda$ is the signal wavelength.. Then a piece of the fibre of length $L$ was cut at the output side of the fibre and the optical power $P_2(\lambda)$ from the fibre was measured again. The attenuation versus wavelength was determined from the general formula $\alpha(dB/m) = 10x \log_{10}[P_2(\lambda)/P_1(\lambda)]/L$.

Fluorescence spectra of the fabricated fibres were also evaluated. The fluorescence measurement was carried out in the contra-propagative configuration using a 350-mW 978-nm laser diode for the excitation of the fibre and a WDM coupler making possible to separate the excitation and fluorescence signals. A short fibre length was used (~14 mm) to avoid distortion by reabsorption and amplified spontaneous emission.

### 2.3. Optimisation of the fabrication process

In order to achieve the targeted core glass composition and designed preform properties, the optimization of different stages of the preparation process was carried out as it is described in following part.

The use of a suitable temperature of deposition of the porous layer is important since the porosity of the layer critically depends on this temperature and the porosity controls dopant incorporation levels in the preform core. This general experience can be specified as follows. High layer porosities achieved at low deposition temperatures allow us to incorporate more dopants but the layer could peel off the tube wall during the solution doping stage, especially when high-viscous soaking solutions are used. On the contrary, the layer porosity and dopants incorporation is reduced at high deposition temperatures due to partial sintering of the deposited layer.

From a set of initial experiments of deposition of porous layers we concluded that for pre-selected flow rates of starting $SiCl_4$ the optimum deposition temperature for the application of a porous layer was between 1250±10°C and 1290±10°C. We have found that an increase of the deposition temperatures above 1300°C makes porous layers more sintered leading to reduction of the dopant incorporation. On the other hand, a decrease of the deposition temperatures below 1210±10°C diminishes the adhesion strength of layers to the glass surface leading to enhanced chance of the layer delaminating during the solution impregnation stage.

To optimize the solution composition, several preforms were fabricated under varying concentrations of the soaking solutions. During this optimization the following concentrations were varied in the following ranges, namely 0.015 to 0.05 M for $LiNO_3$, 0.5 to 1.5 M for $AlCl_3$, 0.015 to 0.05 M for $BaCl_2$, 0.025 to 0.05 M $ZrOCl_2$ and 0.015 to 0.035 M for $ErCl_3$. EMA results were used as the feedback for evaluating the glass compositions obtained from the fabricated preform samples. The optimized composition of the soaking solution was 1.25 M $AlCl_3$, 0.035 M $LiNO_3$, 0.05 M $BaCl_2$, 0.035 M $ZrOCl_2$ and 0.03 M $ErCl_3$. The time for contact of the porous layer and the optimized soaking solution was varied from 30 to 90min. Finally, the time duration of 45min was determined as the optimized soaking period.

After the solution doping stage, oxidation of the raw chemicals was carried out slowly around 900°C in pure oxygen, followed by dehydration around 700°C to avoid evaporation of $Li_2O$ from the porous layer in the form of LiCl due to reaction of $Li_2O$ with of $Cl_2$ used for OH removal.

Sintering of the porous layer with the oxide dopants to a clear glass layer was carried out by a gradual increase of the temperature from 1050°C to 1850°C. This gradual temperature increase is essential to prevent sudden phase separations and reduce the possibility of dopant diffusion from the layer into the surrounding silica layers during the layer sintering.

After collapse of the tube with the deposited layers to the preform the heat-treatment stage took place. This treatment was performed in an electrical furnace which the preform was loaded in. To identify the suitable conditions for the formation of nanocrystalline particles within the preform core, a set of experiments with varied heat-treatment temperature, rate and time span was carried out. In each of these experiments a short section of the fabricated preform was heated at a temperature in a range of 850 to 1100°C and time span from 4 to 7 hours. TEM analysis and electron diffraction pattern of the heat-treated perform core indicates that suitable condition for formation of phase-separated rare-earth doped nanocrystalline particles is : temperature near 1000°C, time span of 5 hours where ramp rate to reach the target temperature was fixed at 7 °C/min. However, this temperature might vary with glass composition and not be considered as definite temperature for all glass compositions.

## 3. Results and discussion

Measured RIPs of the fabricated preforms before and after the thermal treatment showed that their numerical apertures (NA) related to the difference between refractive indexes of the core and the cladding, vary within 0.16±0.02. These values allow us to conclude that there is no essential effect of the heat treatment of the preforms onto their NAs. An example of a representative RIP measured on a preform at different angular projections is shown in Fig 1.

We have found that a RIP of a fibre drawn from the prepared preform is in a good agreement with that of the corresponding preform RIP. The core diameter of the fibre is around 12 μm provided the cladding diameter is 125 μm. It means that the fibre is not single mode at a wavelength of 1550 nm.

The composition of the preform cores prepared with the optimized soaking solution and evaluated from EMA measurements indicates high $Al_2O_3$ incorporation of about 12-14 mol% in all prepared preforms. On the basis of such analysis determined with preciseness better than 0.1 mol% one could expected higher refractive index differences than those shown in

Fig.1. However, even a decrease of NA with an increase of AlCl3 in the soaking solution has been found for highly-doped cores of optical fibers composed of phosphoaluminosilicate glasses prepared by the solution-doping method [23]. This nonliniear dependence of the refractive index on the dopant concentration could be related to mechanical stresses in glasses and by photoelastic effect by density fluctuation caused by formation of crystalline phases. The maximum concentrations for $ZrO_2$ and BaO are 0.9 and 0.2 mol%, respectively. Concentrations of $Er^{3+}$ were also obtained from EMA measurements on the fabricated preforms and compared with those calculated from spectral attenuation curves of the drawn fibres (see an example in Fig. 7). These concentrations vary between 3000 and 4500 mol-ppm in different preforms and the EMA values agree within experimental errors with those measured on the fibres.  Concentrations of Li could not be measured by EMA as lithium atoms are too light for this kind of analysis.

Distributions of $Al_2O_3$, $ZrO_2$ and BaO concentrations in the preform core are presented in Fig 2. The $Al_2O_3$ distribution in Fig.2 shows two distinct regions within the preform core; one which is rich in $Al_2O_3$ (outside the central core region) and second with lower $Al_2O_3$ concentrations (the central dip in $Al_2O_3$ distribution). We presumed that central dip in the $Al_2O_3$ distribution profile formed during the collapsing stage. During collapsing due to high temperature (>2000°C) appreciable volume change occurs as a result of transformation of three phases of $ZrO_2$ (cubic, tetragonal and monoclinic). This transformation associated with change in volume within preform core region and probably forces $Al_2O_3$ away from central core region resulting central dip. Consequently, we could see two distinct phase separated region in Fig. 3a. Transmission electron image was taken from the extracted preform core (see details in Part 2.2.) and it is presented in Fig.3a. It indicates two different regions, namely $Al_2O_3$ rich region (grey colour) and region rich in $ZrO_2$ (black colour). EDX analysis was carried out in different parts under investigation presented in Fig.3a and EDX spot analysis result from the $ZrO_2$ rich part is presented in Fig.3b.

SEM results measured from the central core region of preform in Fig.4 revealed the formation of $ZrO_2$ nanocrystalline particles and characterized by diameters from 20 to 400 nm. However, as one can see from Fig. 4 the most of the nanocrystalline particles have diameters within the 20-80nm range. Although the size of these nanoparticles is-not uniform one can expect that it can be improved by finer optimization of the fabrication steps. To achieve this finer optimization, solution composition specially the ratio between Er/Zr salt will be varied further keeping other solution parameters fixed. Additionally longer annealing period is to be employed to achieve uniform particle size.

The electron diffraction pattern of the particles in the preform is presented in Fig.5 clearly indicating the crystalline nature of the formed particles. SEM results in Fig.6 show that the core-cladding interface is almost uniform in spite of the achieved high $Al_2O_3$ doping. Such uniformity is not usually observed at preforms with high $Al_2O_3$ contents prepared by the standard MCVD and solution-doping techniques [23].

The preform core looks nearly transparent, however, the measured spectral attenuation curve (see Fig. 7) shows high base-line losses which may be either result of phase-separation within the preform core or due to formation of nanocrystalline particles which makes the core material non homogeneous. The presence of particles with diameters large than 50-100 nm in the preform indicate that scattering loss are expected in the fibre. Further work on the perform and fiber production process to lower the background loss to an acceptable value for application such as fiber amplifier.

The fluorescence of fibre drawn from nanoparticle doped preform has been measured in a wavelength region around 1550 nm, under 980-nm excitation at room temperature. The emission curve is shown in Fig. 8 together with the emission from a standard alumino-germano-silicate fibre. Both fluorescence curves have approximately the same shape, except for the shoulder around 1.55 µm: the fibre from the preform containing nanoparticles is slightly lower and some structured shapes are visible, whereas the alumino-silicate fibre has a smooth shoulder. It is out of the scope of this paper to discuss the spectroscopic differences. Though we see that the luminescence from the nanoparticle doped fiber is much broaden compared to that from pure silica (not shown here): we have achieved to modify the close environment of the Er ions, as it is in a reference alumina-germano-silicate erbium-doped fiber, but using a different composition and structuration at a local nanometer scale. Further nanoscale analysis of the fibre structure as well as studies of refined spectroscopic details on these fibres are necessary to analyse whether the nanoparticles are present in the fibre and their effect of the rare-earth ions optical properties.

## 4. Conclusions

Er doped $ZrO_2$-based nanocrystalline phase-separated preforms of optical fibres highly doped with $Al_2O_3$ have been fabricated using the MCVD and solution doping techniques. The prepared preform cores are nearly transparent without defects on the core-cladding interface. The presence of nanocrystalline particles inside the preform core has been observed with non-uniform particle size distributions and diameters within 20-400 nm. Initial results of measurements of fibre shows high base-line optical losses around 1300 nm explained by t strong light scattering on core inhomogeneities such as large nanoparticles or separated phases.

## 5. Acknowledgements

This research was supported by the Czech Science Foundation under contract P/102/10/554. M. Ude and S. Trzésien are acknowledged for fibre geometrical and absorption characterizations of the fibre. Authors are gratefully acknowledging fruitful discussions with other colleagues of IPE and LPMC.

**Figure caption:**

Fig.1: Representative refractive index profile of fabricated preform

Fig. 2: Dopant distributions in the preform core evaluated using EMA

Fig. 3: TEM image of phase-separation in the preform core (a) with spot EDX analysis of $ZrO_2$ rich region (b)

Fig. 4: SEM image showing nanoparticle inside the central core region of the preform

Fig. 5: The electron diffraction pattern of particle formed indicating crystalline nature

Fig. 6: SEM image of the preform core and the core-cladding interface

Fig. 7: Spectral losses of the fabricated Er-doped fibre

Fig. 8: Normalized fluorescence curves from Er-doped fibres excited at 980 nm at room temperature. Thick solid line: fibre drawn from the nanocrystallite containing preform. Thin dashed line: standard alumino-germano-silicate fibre (see text)

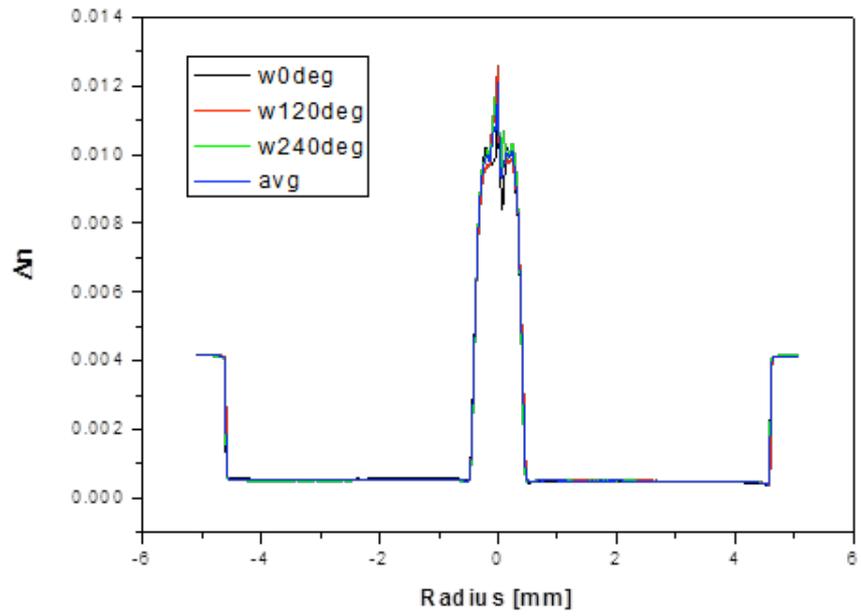

Fig.1: Representative refractive index profile of fabricated preform

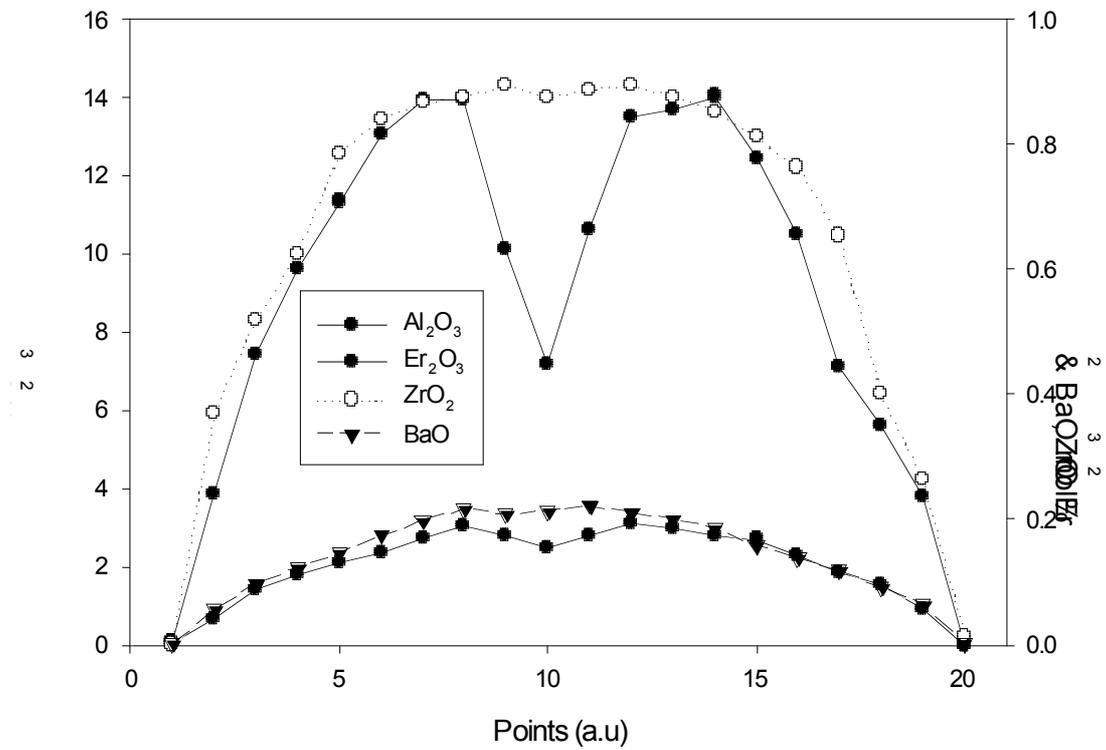

Fig. 2: Dopant distributions in the preform core evaluated using EMA

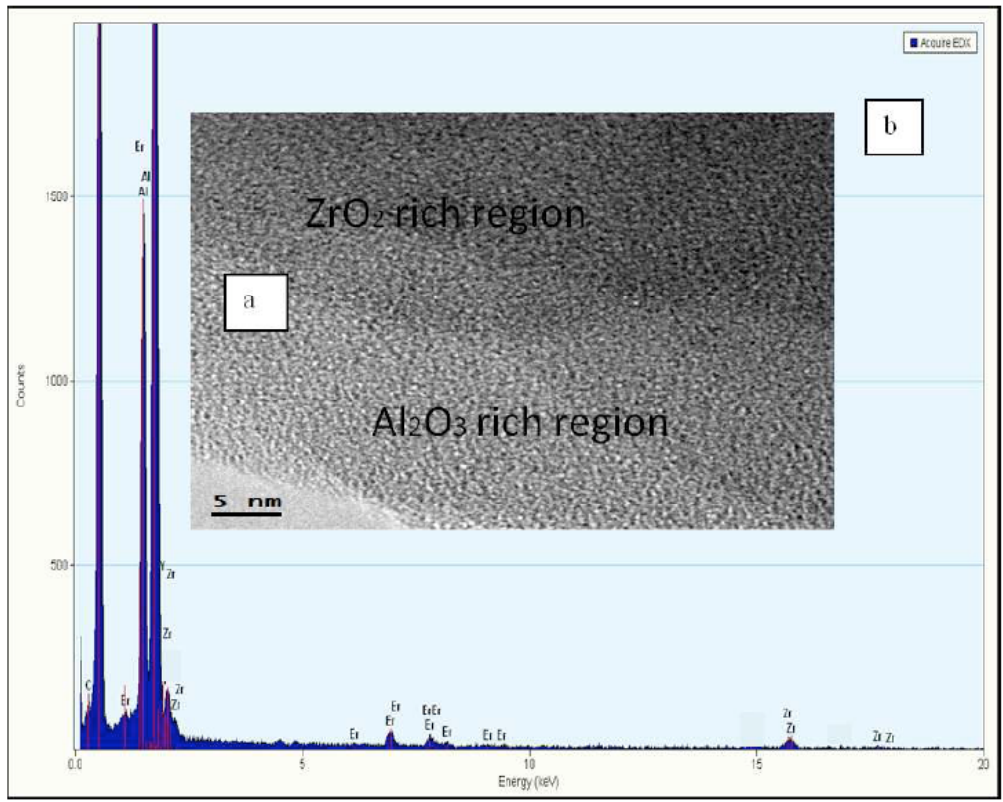

Fig. 3: TEM image of phase-separation in the preform core (a) with spot EDX

analysis of $ZrO_2$ rich region (b)

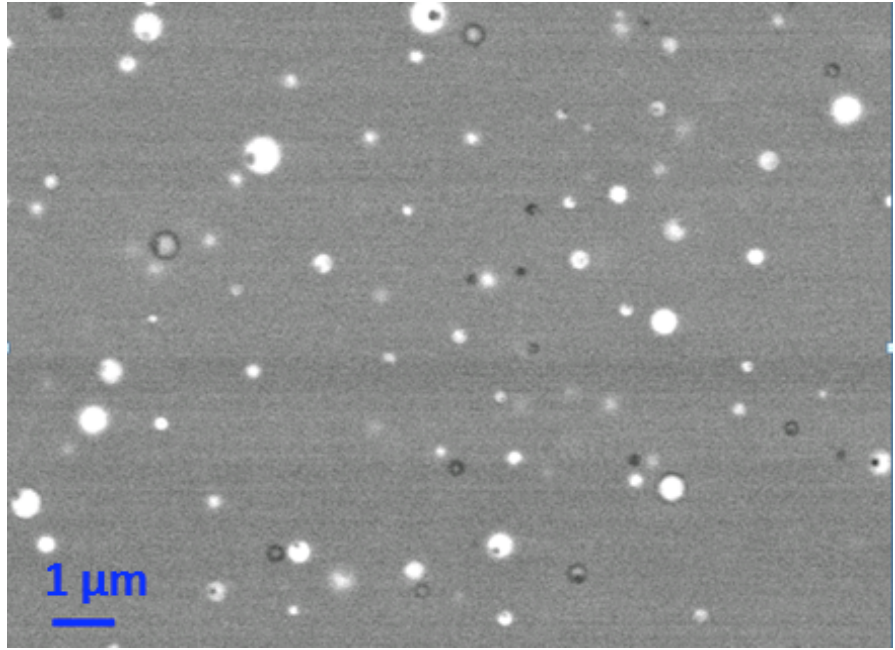

Fig. 4: SEM image showing nanoparticle inside the central core region of the preform

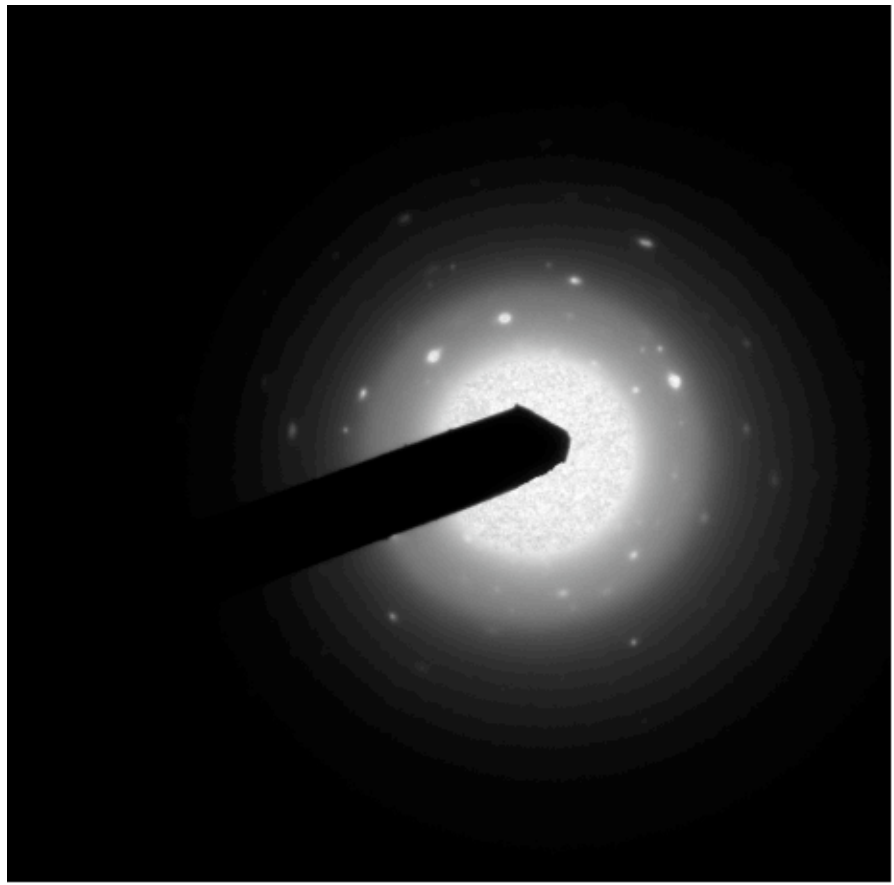

Fig. 5: The electron diffraction pattern of particle formed indicating crystalline nature

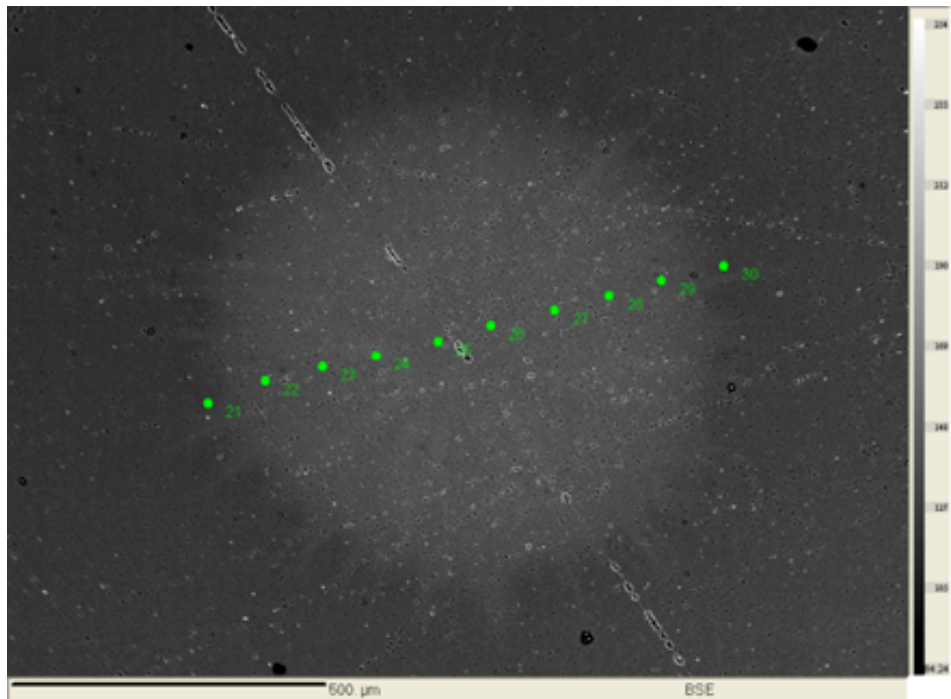

Fig. 6: SEM image of the preform core and the core-cladding interface

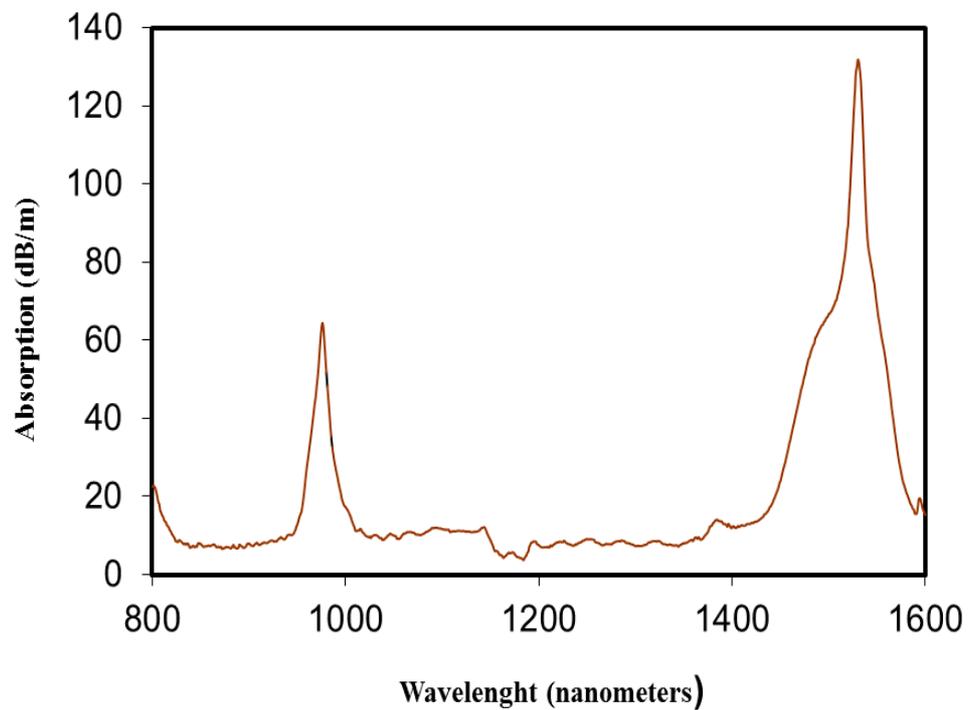

Fig. 7: Spectral losses of the fabricated Er-doped fibre

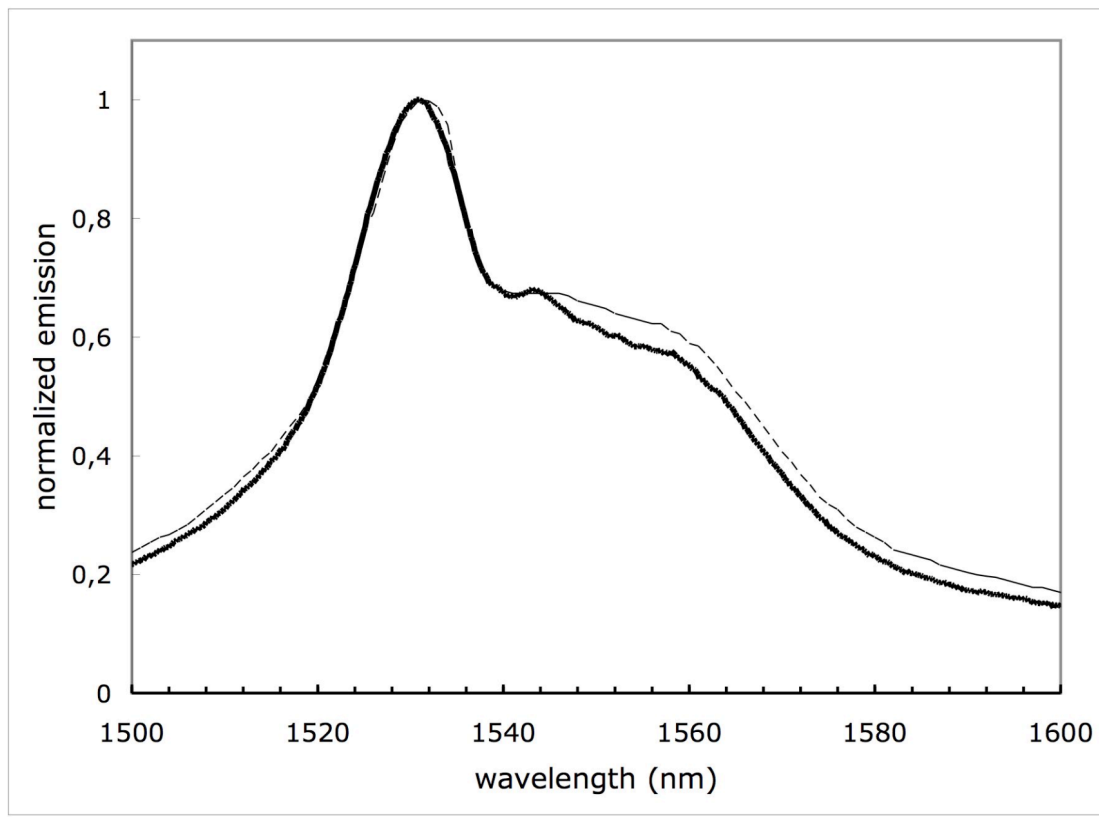

Fig. 8: Normalized fluorescence curves from Er-doped fibres excited at 980 nm at room temperature. Thick solid line: fibre drawn from the nanocrystallite containing preform. Thin dashed line: standard alumino-germano-silicate fibre (see text)